# Vacancy Induced Energy Band Gap Changes of Semiconducting Zigzag Single Walled Carbon Nanotubes

Gulay DERELI[1], Onder EYECIOGLU[2], Banu SÜNGÜ MISIRLIOĞLU[3]

[1]*Department of Physics (Emeritus), Faculty of Arts and Sciences, Yildiz Technical University, Istanbul, 34220, Turkey*

[2]*Department of Computer Engineering, Faculty of Engineering and Architecture, Nisantasi University, Istanbul, 34406, Turkey*

[3]*Department of Physics, Faculty of Arts and Sciences, Yildiz Technical University, Istanbul, 34220, Turkey*

onder.eyecioglu@nisantasi.edu.tr

*Abstract*— — In this work, we have examined how the multi-vacancy defects induced in the horizontal direction change the energetics and the electronic structure of semiconducting Single-Walled Carbon Nanotubes (SWCNTs). The electronic structure of SWCNTs is computed for each deformed configuration by means of real space, Order(N) Tight Binding Molecular Dynamic (O(N) TBMD) simulations. Energy band gap is obtained in real space through the behavior of electronic density of states (eDOS) near the Fermi level. Vacancies can effectively change the energetics and hence the electronic structure of SWCNTs. In this study, we choose three different kinds of semiconducting zigzag SWCNTs and determine the band gap modifications. We have selected (12,0), (13,0) and (14,0) zigzag SWCNTs according to n (mod 3) = 0, n (mod 3) = 1 and n (mod 3) = 2 classification. (12,0) SWCNT is metallic in its pristine state. The application of vacancies opens the electronic band gap and it goes up to 0.13 eV for a di-vacancy defected tube. On the other hand (13,0) and (14,0) SWCNTs are semiconductors with energy band gap values of 0.44 eV and 0.55 eV in their pristine state, respectively. Their energy band gap values decrease to 0.07 eV and 0.09 eV when mono-vacancy defects are induced in their horizontal directions. Then the di-vacancy defects open the band gap again. So in both cases, the semiconducting-metallic - semiconducting transitions occur. It is also shown that the band gap modification exhibits irreversible characteristics, which means that band gap values of the nanotubes do not reach their pristine values with increasing number of vacancies.

## I. INTRODUCTION

Soon after their discovery by Iijima in the early 1990s [1], carbon nanotubes have attracted considerable attention due to their extremely small dimensions, mechanical strength, as well as, their elasticity and adjustable electronic properties. In spite of their small dimensions, adjustable electronic characteristics for a metallic or semiconducting conductivity are structure dependent. These remarkable characteristics led to several nanotechnology applications such as gas sensors, dielectric devices, nanoelectronic devices and emitters [2,3]. However, experimental studies also showed that structural defects are often present in nanotubes [4,5]. It is well-known that defects affect the electronic properties of SWCNTs in a drastic way. Several kinds of defects may occur on carbon nanotubes. Among them, vacancy defects attract considerable attention. They may occur during their growth, purification or device application processes. Furthermore, vacancies can also be created in the laboratory through electron or ion irradiations [6,7] by removing individual carbon atoms from SWCNTs. Following up these results, some of the potential applications of vacancy defected SWCNTs such as chemical sensors [8], catalysts for dissociation of water [9] and field effect transistors [10] are given in literature. In this context, it seems meaningful to study the electronic properties of vacancy defected SWCNTs. Theoretical calculations have shown that the vacancy defects can substantially modify the electronic properties of zigzag SWCNTs [10–20]. In Refs. [10,11], the electronic band gap modification of mono-vacancy defected SWCNTs are studied. The effect of multi-vacancies on the electrical properties of SWCNTs are also performed by Density Functional Theory (DFT) based on first principle calculations [12–19] and some empirical Tight Binding (TB) models [20]. These results are given with respect to the number and configuration of vacancy defects. It is shown that conductance variation of (12,0) zigzag carbon nanotubes, which leads to n(mod 3) = 0 classification, is not a monotonic function of vacancies and it relates to the reconstruction around the defects [12]. In Ref. [13], different types of defect configurations such as a vacancy-adatom complex, pentagon-octagon-pentagon topological defects are studied and localized states in the band gap of (17,0) zigzag carbon nanotubes are concluded. In Ref. [14] according to n(mod 3) = 1 classification, the effect of multi-vacancies on the electronic structure of (10,0) semiconducting carbon nanotube is investigated and di-vacancy is found to have a significant effect on the band gap of the semiconducting nanotubes. Additionally, single and double vacancy defected (10,0) semiconducting and (5,5) metallic carbon nanotubes are studied by some of the other researchers and they have concluded that introduction of vacancy on the carbon nanotubes decreases the band gap of semiconducting nanotubes, increases the band gap of

This work was supported in part by Yildiz Technical University Research Fund Project No: 2009-01-01-KAP01

metallic nanotubes. In Ref. [16] according to n(mod 3) = 2 classification, (8,0) and (14,0) semiconducting carbon nanotubes are investigated. It is concluded that with respect to the increasing number of vacancies, the band gap value of the tubes decreases and as a result defected tubes transform into metallic conductors. Some of the other researchers also studied the band gap modification of armchair SWCNTs that show a metallic behavior in their pristine state [17-19]. According to their results, the band structure of an (5,5) armchair SWCNT is very sensitive to vacancies and the influence of them extends to the entire band structure. On the other hand, an empirical Tight Binding (TB) method [20] is performed in order to investigate the effects of vacancies on the electronic structure of both (20,0) semiconducting and (12,12) metallic carbon nanotubes. It is shown that, the occurrence of vacancies induces an increase of density of high-energy occupied states and the band gap modification is obtained. In the lights of these results, it is shown that vacancy defects have a crucial effect on the electronic behavior of zigzag carbon nanotubes. Even though previous results are beginning to yield a general picture for the conducting behavior of various vacancy defected tubes, some critical conclusions are still unresolved. It is seen that n(mod 3) classification is not taken into account. From this point of view, it is necessary to give more systematic investigations according to the n(mod 3) = 0, n(mod 3) = 1 and n(mod 3) = 2 classification for (n,0) zigzag SWCNTs. In our work for the first time, all of the three classifications are taken into account with respect to the increasing number of vacancies and energy band gap modifications are clarified. In all these cases, multi-vacancy defects are induced in the horizontal direction. The energetics and electronic structure simulations of the three types of zigzag SWCNTs are performed. Energy band gap is displayed in real space through the calculations of electronic density of states (eDOS) near the Fermi level. The electronic band gap modification of originally metallic or semiconducting zigzag SWCNTs are discussed with respect to the increasing number of vacancies. Zigzag SWCNTs with n (mod 3) = 0 are metallic while the others (i.e. n (mod 3) = 1 and n (mod 3) = 2) are semiconducting in pristine (undeformed) conditions. The numbers of the vacancies that cause the first metal-semiconductor and semiconductor-metal transitions in each zigzag SWCNTs are found. The Fermi energy levels, electronic density of states (eDos) and energy band gaps of the selected zigzag SWCNTs are discussed with respect to the increasing number of vacancies. The electronic structure of SWCNTs have been computed for each deformed configuration by means of real space, O(N) parallel TBMD algorithms which is successfully applied to the SWCNTs by G. Dereli [21–23]. O(N) TBMD calculates the band structure energy in real space and makes the approximation that only the local environment contributes to the bonding, and hence the band energy of each atom. The running time scales are efficiently linear with respect to the number of atoms. The key idea of this approach is to provide a description of a large system in terms of contributions coming from its subsystems. The system is split into subsystems and each subsystem is solved independently. Each subsystem comes with its own Hamiltonian matrix, constructed relative to the corresponding buffer. The details of our simulation methods is also given in Section II. Additionally, according to our previous works in literature [24–27], it has been shown that our simulation method works very well in identifying the physical, electronic and mechanical properties of SWCNTs under various conditions such as temperature, strain etc.. In Ref. [24,25], the mechanical properties of armchair and zigzag SWCNTs are studied up to the high temperatures. Additionally, in Ref. [26] the energetics and thermal stability of carbon nanotubes are also concluded up to the high temperatures. In our work of [27], we have examined successfully the band gap modification of zigzag carbon nanotubes with respect to the tensile strain. As a results, it is apparent that our real space O(N) parallelization technique is a sufficient tool for analyzing the electronic behavior of vacancy defected SWCNTs. We aimed to reach one step ahead from the studies in literature. In this work, we have selected (12,0), (13,0) and (14,0) zigzag SWCNTs according to n (mod 3) = 0, n (mod 3) = 1 and n (mod 3) = 2 classification. The tubes are optimized approximately in the same tube length of 1.99 nm so they consist of 240, 260, 280 carbon atoms, respectively. The simulations are performed at 300K so that the effects of temperature may be neglected and only the brittle SWCNTs [24,26] are taken into account.

According to the work flow chart of this paper, the basis of the O(N) TBMD technique that we used and the electronical calculations on the energy band gaps of the optimized pristine/vacancy defected SWCNTs are explained in Section II. Also in Section III, the vacancies that are applied in the horizontal direction of the tubes are clarified and the results such as total energy, electronic density of states (eDOS) and energy band gap modifications are given with respect to the increasing number of defects. In Section IV, our results are concluded.

## II. METHOD

The O(N) parallel TBMD simulation technique which was successfully applied to SWCNTs by G.Dereli [21–23], is used in order to study the energetics and electronic properties of vacancy defected (12,0), (13,0) and (14,0) SWCNTs. The details of our electronical calculations are summarized below:

### II.1. TIGHT BINDING MOLECULAR DYNAMIC (TBMD) SIMULATION METHOD

In the classical TBMD, the electronic structure of a system may be calculated by the TBMD Hamiltonian

$$H_{TBMD} = \sum_\alpha \frac{P_\alpha}{2m_\alpha} + \sum_n \varepsilon_n f(\varepsilon_n, T) + U_{rep} \quad (1)$$

so that the quantum mechanical many-body nature is taken into account [28,29]. The forces needed to move the atoms are evaluated from the above TBMD Hamiltonian as

$$\vec{F}_{tot,\alpha} = -\sum_n \langle \Psi_n | \frac{\partial H}{\partial \vec{r}_\alpha} | \Psi_n \rangle f(\varepsilon_n, T) - \frac{\partial U_{rep}}{\partial \vec{r}_\alpha} \quad (2)$$

The second term on the right hand side of Eq. (2) is a repulsive force given analytically as a function of the interatomic distances [30]. On the other hand the first term of the Eq. (2) is the Hellmann-Feynman contribution to the total force,

$$\sum_n \langle \Psi_n | \frac{\partial H}{\partial \vec{r_\alpha}} | \Psi_n \rangle f(\varepsilon_n, T) = \\ -2\sum_n f(\varepsilon_n, T) \sum_{l\gamma} \sum_{l'\beta} C^n_{l'\beta} \frac{\partial H_{l',\beta,l,\gamma}(\vec{r_{\beta\gamma}})}{\partial \vec{r_\alpha}} C^n_{l\gamma} \quad (3)$$

Furthermore, the total energy of a system of ion cores and valance electrons can be written as:

$$E_{tot} = 2\sum_n f(\varepsilon_\alpha, T)\varepsilon_n + U_{ii} - U_{ee} = E_{bs} + U_{rep} \quad (4)$$

where $f(\varepsilon_n,T)$ is the Fermi-Dirac distribution function. The sum of all the single particle energies is commonly called the band structure energy $E_{bs}$.

In order to calculate the band structure energy and Hellman-Feynman forces, we need the full spectrum of eigenvalues $\varepsilon_n$ determined by solving the secular equation

$$\sum_{l'\beta}\left( \langle \varphi_{l'\beta} | h | \varphi_{l\alpha} \rangle - \varepsilon_n \delta_{ll'}\delta_{\alpha\beta} \right) C^n_{l'\alpha} = 0 \quad (5)$$

and the corresponding eigenvectors should also be found.

II.2. ORDER N TIGHT BINDING MOLECULAR DYNAMICS (O(N) TBMD)

The TBMD Hamiltonian matrix defined in Eq. 5 must be diagonalized at every time step of the simulation. Standard diagonalization of the TB matrix requires computation times in cubic scaling with respect to the number of atoms (O(N3)) [21–23]. Therefore, it is very difficult to simulate very large systems (i.e N>1000). By reducing the complexity of the algorithm to the linear complexity named as O(N) methods, it is possible to simulate larger systems in long simulation time. In the O(N) methods, band structure energy is calculated in real space and only local contributions are taken into the account [30-33].

In this study, "Divide and Conquer" approach (DAC) is used as the O(N) method. In this approach, a physical large system is divided into subsystems and the charge distribution of whole system is defined as the sum of all contributions that come from subsystems. The band structure energy of the system depending on eigenvalue of one electron Hamiltonian ($\epsilon_i$) and Fermi – Dirac distribution ($F(\epsilon_i)$) is formulated as [32]

$$E_{bs} = 2\sum_i^{Nb} F(\epsilon_i) \epsilon_i \quad (6)$$

where $F(\epsilon_i)$ is Fermi- Dirac distribution.

$$F(\epsilon_i) = \frac{1}{1+e^{\left(\frac{\epsilon_i - \mu}{kT}\right)}} \quad (7)$$

By calculating the sum of the electron densities of each subsystem, the solution of the whole system is obtained. Each sub system is defined by local base functions instead of atomic orbitals. Secular equation of subsystems is written as;

$$H^\alpha C_i^\alpha = S^\alpha C_i^\alpha \epsilon_i^\alpha \quad (8)$$

II.3. PARALLELIZATION ALGORITHMS

O(N) algorithms are suitable algorithms for parallelization. By applying parallel algorithms to the O(N) methods, the better performance can be obtained for simulation run time.

The simulation program used in this study is written in Fortran 77 programming language by G.Dereli et al. [21,22]. In this program, "Single Instruction Multiple Data (SIMD) taxonomy" is used for process parallelization. To keep communication time to minimum only band structure energy and force calculations are parallelized. PVM (Parallel Virtual Machine) software is used for massage passing model of heterogeneous UNIX network. Replicated data (RD) and data decomposition (DD) algorithms are used for data parallelization. RD algorithm is used to send atomic coordinates to all computation nodes. The DD algorithm divides a simulation box into equal-sized boxes and uses them to assign as sub-cubical box of each computation node. The value of cube size that defines the size of a sub-system is taken as distance between two cross-sectional layers as seen in Fig.1. This definition provides the same number of interacting neighbor atoms (buffer) for each subsystem. The periodic boundary condition (PBC) is only applied in the uniaxial direction of SWCNT. So, the SWCNT acts as an infinite length tube.

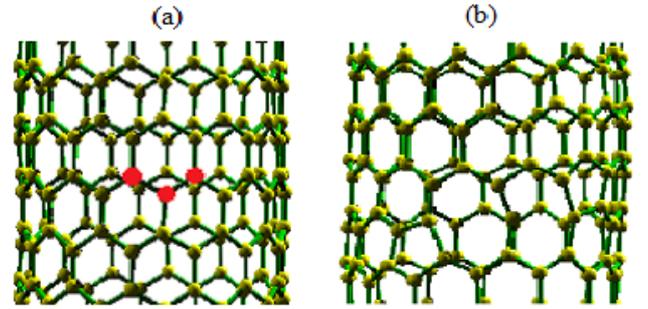

Figure 1. Optimized structures of (a) pristine, (b) vacancy-defected SWCNTs (trio-vacancy is given as an example.)

The simulations are realized in a Linux High Performance Computing (HPC) cluster which has "star" network topology.

II.4. IDENTIFYING VACANCY DEFORMATIONS

In this study, O (N) TBMD simulation code which is written by G.Dereli et.al [21,22] was modified in a manner which can create one or more vacancy on equatorial plane of SWCNT. Vacancies are placed perpendicular to the CNT axis. The flowchart of the vacancy modification of the program is given in Fig. 2.

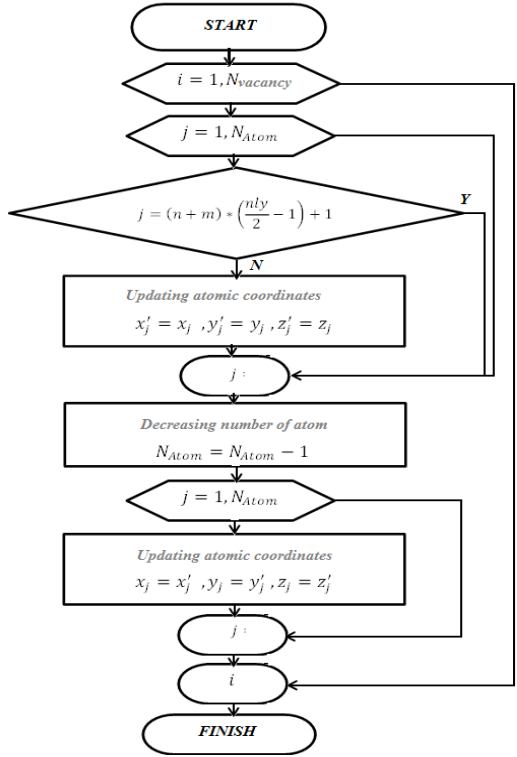

Figure 2. The process flowchart of vacancy subroutine

### II.5. ENERGY BAND GAP CALCULATIONS

The key variable is the energy band gap to define the electronic characteristic of the SWCNTs. The energy band gaps of optimized pristine and vacancy defected SWCNTs are determined from real space electronic density of states (eDOS) which is obtained from the general formula

$$\rho(\varepsilon) = \frac{dN(\varepsilon)}{d\varepsilon} = \frac{N(\varepsilon - E) - N(E)}{E} \quad (9)$$

where N is number of electrons in the system and

$$N(\varepsilon) = \sum_{i=1}^{N_{CELL}} \sum_{}^{4N_{at}} \frac{2}{1 + f\left((E - \varepsilon_i)/k_B T\right)} * \sum_{j=1}^{4N_{at}} |H(j,i)|^2 \quad (10)$$

The existence of populated electronic states near the Fermi energy level on the eDOS (Eq. 6) determinates the energy band gap. During our simulations, we first optimized a pristine SWCNT and let it reach its equilibrium state in 3000 MD steps of simulation time (each time step is chosen to correspond to 1fs). A greater number of MD steps were tested and energy fluctuations are found to remain unchanged. At the following stage, a certain number of vacancies are created on the tube structure and then the system is allowed to take the next 3000 MD steps to reach its new modified equilibrium state. The process is repeated as many times as desired.

### III. RESULTS AND DISCUSSIONS

The vacancy defected (12,0), (13,0) and (14,0) zigzag SWCNTs are studied in order to overcome the difficulties in characterizing multi-vacancy defects induced by the horizontal direction changes; the energetics and the electronic structures of semiconducting Single-Walled Carbon Nanotubes (SWCNTs).

The applied simulation procedure in which we have obtained the results is obtained as follows;

i. The initial atomic coordinates and velocity values that describe the undeformed (pristine) SWCNT are defined.
ii. Kinetic energy and the potential energy values that include ion – ion and ion-electron interactions are calculated with Eq. 4.
iii. New atomic coordinates and velocities are updated with Eq. 1 and 2 according to Velocity Verlet integration algorithm [34].
iv. A pristine zigzag (n,0) SWCNT is optimized and then certain amount of vacancies are generated on the tube structure. Following this procedure, the effect of vacancy defects on the total energy per atom (Etot) and electronic density of states (eDOS) are simulated at 300K.
v. The energy band gap values are calculated from the behavior of electronic density of states near the Fermi energy level.

The vacancies are created in the horizontal direction, along the equator of the zigzag SWCNT is given in Fig. 1. According to Figure 1a and b, a close up view of our optimized pristine structure and the corresponding vacancy defected one are given, respectively. The red atoms in Figure 1a are labeled for the vacancy defects that are induced in Figure 1b. According to Figure 1b, it is seen that, our simulation procedure allows the carbon structure to reach its structural stability and let the reconstruction of carbon-carbon bonds around vacancies, which led to octagons- pentagon pairs on the structure.

Our simulation procedures for (12,0), (13,0) and (14,0) zigzag SWCNTs are given in Fig.3 (a), (b) and (c), respectively. In the first 3ps of simulation time, pristine tubes are allowed to reach their structural stability. During the next 3ps of simulation time, defects are created along the horizontal direction of the nanotubes with the mono- di- trio- tetra-quintet vacancies and the structures are allowed to reach their new equilibrium state. As shown in total energy fluctuations in Fig.3, vacancy defected (12,0), (13,0) and (14,0) zigzag SWCNTs sustain their stabilities for the introduced number of vacancies. The total energy values that are calculated from Fig.3 are figured as functions of the number of the vacancy numbers and shown in Fig. 4. This figure shows that the total energy per atom increases linearly with increasing number of vacancies.

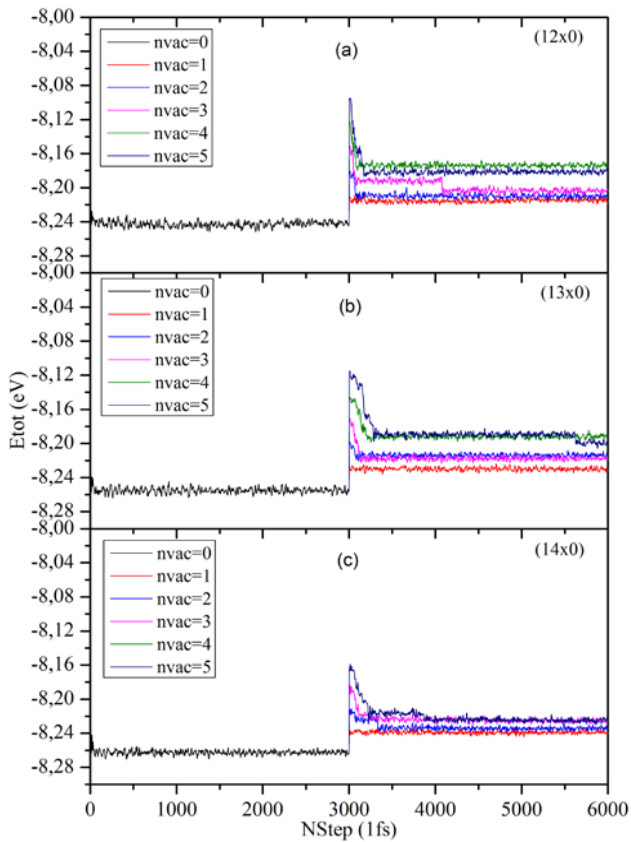

Figure 3. Variation of the total energy per atom (eV) with the increasing number of vacancy. First 3000 MD steps give the energy of pristine SWCNTs (without vacany), next 3000 MD steps give the energy of defected SWCNTs.

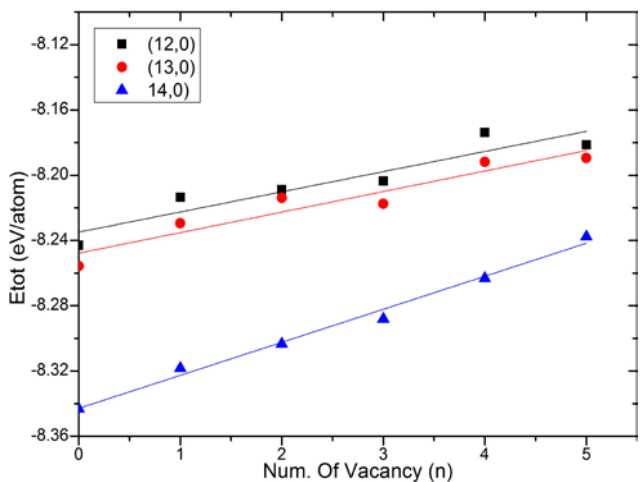

Figure 4. Total energy (eV/Atom) versus number of vacancy deviation for zigzag (12,0) , (13,0) and (14,0) SWCNT.

Next, we investigate the electronic density of states (eDOS) of the nanotubes. eDOS of (12,0), (13,0) and(14,0) zigzag SWCNTs are calculated from Eq. (6) and plotted in Figures 5, 6 and 7, respectively. eDOS graphs are normalized by moving the Fermi energy level to zero along the x axis. The existence of electronic states near the Fermi energy level through the eDOS graphs defines the energy band gap (Eg) of SWCNTs. Figure 5 shows the changes in the eDOS of (12,0) SWCNT with the increasing number of vacancy defects. It is shown that pristine (12,0) tube (nvac=0) has no band gap, indicating metallic behavior in its pristine state. When the vacancy defects are introduced in the horizontal direction, a band gap appears and it opens up. The maximum band gap is seen for the di-vacancy defected tube (nvac=2). For tri-vacancy defected (12,0) SWCNT, the band gap closes and it approximately sustains its width with respect to the increasing number of vacancies (nvac=4,5). Figure 6 gives the change in the eDOS of (13, 0) SWCNT with increasing number of vacancy defects. It is shown that the (13, 0) SWCNT has a band gap in the pristine state which closes for the mono-vacancy defected tube, thus indicating a semiconducting-metallic transition. On the other hand, when di-vacancies are introduced in a (13,0) SWCNT, band gap opens that means that tube recovers its semiconducting behavior, after which it approximately sustains its width with respect to the increasing number of vacancies (nvac=3,4,5). Figures 7 shows the eDOS behavior of a (14,0) SWCNT, with increasing number of vacancy defects. It is shown that (14,0) SWCNT has a band gap in the pristine state that closes for mono-vacancy defected (14,0) SWCNT. However, the band gap opens again for di-vacancy defected (14,0) SWCNT and then it approximately sustains its width with respect to the increasing number of vacancies (nvac=3,4,5). This behavior is similar to that of (13,0) SWCNT.

The energy band gaps of (12,0), (13,0) and (14,0) zigzag SWCNTs are obtained from the existence of electronic states near the Fermi energy level on the eDOS graphs and they are given respectively, as a function of vacancy in Figure 8. As shown in this figure, the band gap of (12,0) nanotube is 0.01 eV in the pristine state. When a mono-vacancy is introduced in the horizontal direction the band gap opens and reaches 0.13 eV for the di-vacancy defected tube. However, for trio-vacancy defected (12,0) nanotube, the band gap value decreases and sustains its width approximately up to the quintet-vacancies. These are consistent with the theoretical calculations of Refs [12]. They have shown that di-vacancy defected (12,0) tube has lower conductance than the tetra-vacancy defected one. In this point of view, we have also presented the energy band gap modification with respect to the number of vacancy in order to clarify the opening band gap value of di-vacancy defected (12,0) nanotube.

In Figure 8, it is also given that (13,0) and (14,0) SWCNTs are semiconductors having energy band gap values of 0.44 eV and 0.55 eV, respectively, in their pristine state. Energy band gap values decrease to 0.07 eV and 0.09 eV, respectively, when mono-vacancy defects are introduced in their horizontal direction. Therefore in both cases, the energy band gap closes and semiconducting – metallic transitions are observed. On the other hand, di-vacancy defects open the band gap again and the gap sustains its width around 0.2 eV up to the quintet-vacancies. In the work of Ref. [14], a significant di-vacancy modulation is obtained for n(mod 3=1) classification. In Ref. [20] the energy band gap value of (20,0) SWCNT (n(mod 3=2)) is given as 0.39 eV in the pristine state and 0.18 eV for the multi-vacancy defected one. However, the semiconducting-metallic transitions are not mentioned effectively. In our results, we have shown that an effective band gap modulation is dominant between mono-di-tetra vacancy deviations which led to a significant metallic-semiconducting-metallic transition for both n(mod 3=1) and n(mod 3=2) classified zigzag SWCNTs. Additionally, results of Ref. 16 are also in good agreement with ours for n(mod 3=1) type of zigzag carbon nanotubes.

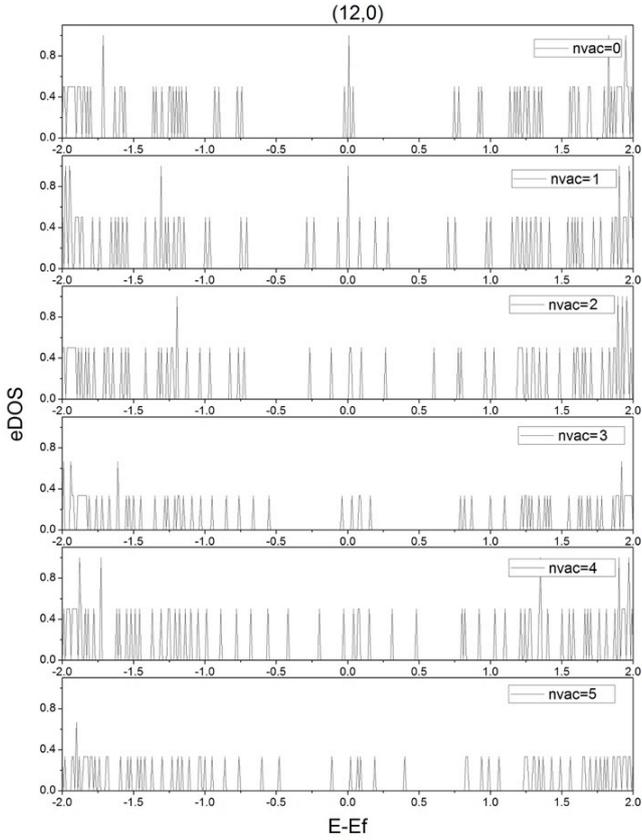

Figure 5. eDOS structures of pristine and vacancy defected (12,0) SWCNT.

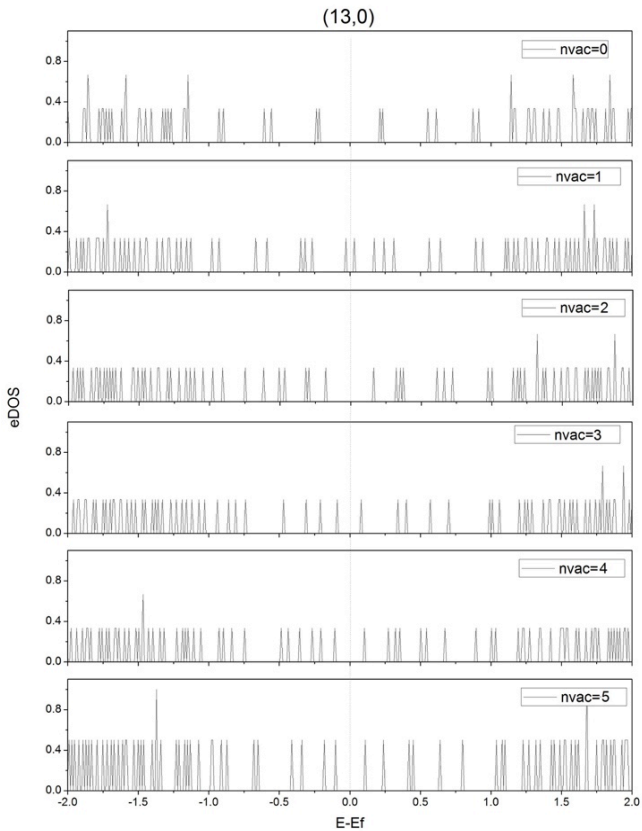

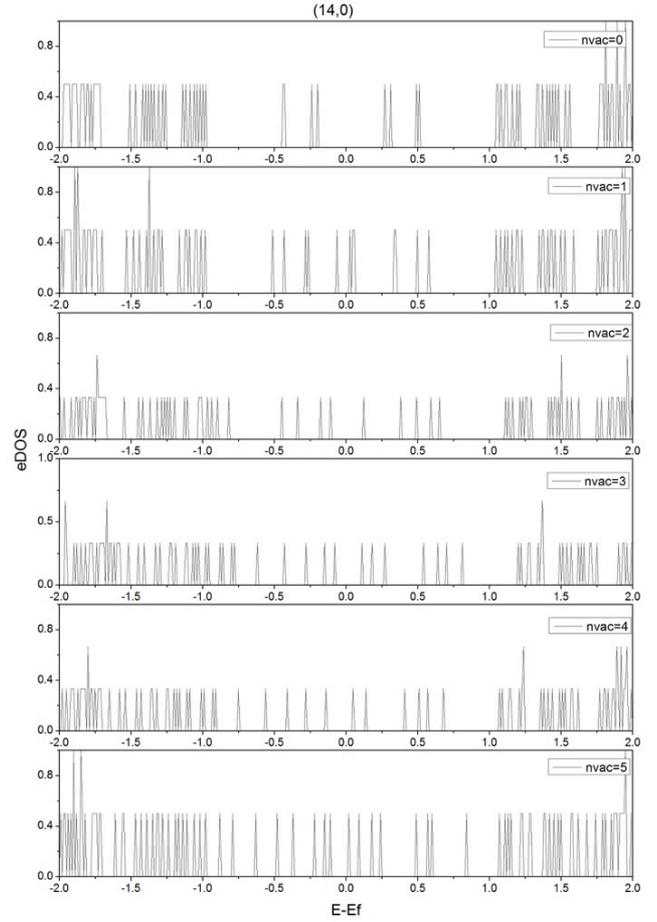

Figure 6. eDOS structures of pristine and vacancy defected (13,0) SWCNT.

Figure 7 eDOS structures of pristine and vacancy defected (14,0) SWCNT

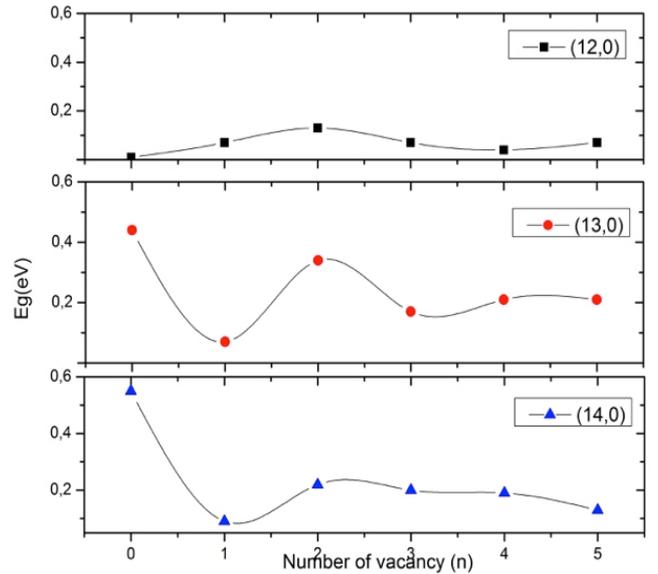

Figure 8. Energy band gap (eV) modification of (12,0), (13,0) and (14,0) zigzag SWCNTs as a function of vacancy

## IV. CONCLUSION

This study provides useful insights for understanding the electronic behavior of deformed zigzag SWCNTs. We have used O(N) parallel TBMD simulation method [21–23] in order to investigate the energetics and the electronic properties of zigzag SWCNTs with respect to the increasing number of vacancies. The advantage of our technique is based on the fact that the calculations are done in real space. It makes the approximation that only local environment contributes to the bonding, and hence bond energy, of each atom. Moreover, this study shows that O(N) TBMD simulation method is an efficient tool for clarifying the tensile behavior and electronic structure of SWCNTs which are important for designing the future nanotube based electronic and optoelectronic devices. The effect of the multi-vacancy defects on the stability, total energy, eDOS structure and the energy band gap of the zigzag SWCNTs are investigated. The zigzag SWCNTs (n,0) are classified into three groups: n(mod 3) = 0 , n(mod 3) = 1 and n(mod 3=2). We have chosen according to these classification scheme, (12,0), (13,0) and (14,0) zigzag SWCNTs, respectively.

Our studies showed that vacancy defects created in the horizontal direction, along the equator of SWCNTs can effectively change the energetics and hence the electronic structure of SWCNTs. The total energy increases linearly with increasing number of vacancies and the tube structure sustains its structural stability up to the quintet-vacancy defected SWCNTs. The energy band gaps of (12,0), (13,0) and (14,0) zigzag SWCNTs are obtained from the existence of electronic states near the Fermi energy level on the eDOS graphs. It is shown that eDOSes are enhanced as the vacancy defects are introduced on the SWCNTs, depending on the classification of the tubes. Zigzag SWCNTs that are represented as n(mod 3) = 0 are metallic while the others (i.e. n(mod 3) = 1 and n(mod 3) = 2) are semiconducting in their pristine states. The energy band gap of n(mod 3) = 0 zigzag SWCNT increases and a relatively decreasing conducting effect is seen for di-vacany defected (12,0) tube. On the other hand, for n(mod 3) = 1 and n(mod 3) = 2 types of zigzag SWCNTs, the energy band gaps decrease effectively for the mono-vacancy defected ones and then increases again for di-vacancy defected tubes. So a dominant semiconducting-metallic-semiconducting transitions are concluded. With the increasing number of vacancies the tube recovers its semiconducting behavior with a narrow band gap. The vacancy dependant band gap modification implies irreversible characteristics, which means that the energy band gap values of the nanotubes do not reach their pristine value with increasing number of vacancies.

In summary our results clarify the energy band gap modifications for the multi-vacancy defected zigzag SWCNTs. The most drastic band gap modification is obtained for the mono-vacancy defected (n(mod 3) = 1) and (n(mod3)= 2) zigzag SWCNTs as a semiconducting-metallic transition in their conductivity.  For the remaining type of zigzag SWCNTs (n(mod 3) = 0), the multi-vacancy defects create a narrower  band gap, however, the tubes recover their semi-metallic behavior. In view of our results, the adjustable band gap modification attracts attention for the performance of nanotube based electronic devices, which sheds light on the future potential applications of the defected zigzag SWCNTs.

ACKNOWLEDGMENT

The research reported here is supported through the Yildiz Technical University Research Fund Project No: 2009-01-01-KAP01.Simulations are performed at the "Carbon Nanotubes Simulation Laboratory" at the Department of Physics, Yildiz Technical University, Istanbul, Turkey.